\documentclass[a4paper]{article}

\newcommand{\dessin}[4]{
\begin{figure}[!h]
\begin{center}
\resizebox{#3cm}{!}{\includegraphics*{#2}}
\caption{#4}
\label{#1}
\end{center}
\end{figure}}

\usepackage{RR}
\usepackage{hyperref}
\usepackage{color}

\RRNo{5915}

\RRdate{Octobre 2006}
\RRauthor{
Cédric Adjih
\and
Daniel Augot
  \and
Raghav Bhaskar
\and
Saadi Boudjit
\and
Paul Mühlethaler
\and

}
\authorhead{Adjih et al}
\RRtitle{AGDH (Asymetric Group Diffie Hellman),
un protocole de mise en accord de clé efficace pour les réseaux Ad Hoc}
\RRetitle{AGDH (Asymetric Group Diffie Hellman) 
An Efficient and Dynamic Group Key Agreement Protocol for Ad Hoc Networks}
\titlehead{AGDH}
\RRresume{ Les problèmes de confidentialité, d'intégrité et
  d'authentification sont de plus en plus prévalents dans les réseaux
  Ad Hoc, mais aussi dans les réseaux fixes filaires. Une approche à
  ces problèmes est d'utiliser la cryptographie symétrique (ou à clé
  secrète), reposant sur une clé partagée par tous les membres du
  réseau. Mais étblir et maintenir une telle clé, dite de session, est
  un problème non trivial. Nous montrons que les protocoles de mise en
  accord de clé de groupe (GKAs : Group Key Agreement protocols) sont
  bien adaptés pour établir et maintenir de telles clés de session
  dans les réseaux dynamiques. Nous considérons un protocole déjà
  établi, qui est robuste aux pertes de onnectivité, et nous
  envisageons tous les problèmes relatifs au bon fonctionnement de ce
  protocole dans les réseaux Ad Hoc. Nous donnons des détails
  d'implémentation, des paramètres réseaux, ce qui permet de réduire
  considérablement la charge calculatoire liée à l'emploi de la clé
  publique dans de tels réseaux.
  }

\RRabstract{Confidentiality, integrity and authentication are more
  relevant issues in Ad hoc networks than in wired fixed networks. One
  way to address these issues is the use of symmetric key
  cryptography, relying on a secret key shared by all members of the
  network. But establishing and maintaining such a key (also called
  the session key) is a non-trivial problem. We show that Group Key
  Agreement (GKA) protocols are suitable for establishing and
  maintaining such a session key in these dynamic networks. We take an
  existing GKA protocol, which is robust to connectivity losses and
  discuss all the issues for the good functioning of this protocol in
  Ad hoc networks. We give implementation details and network
  parameters, which significantly reduce the computational burden of
  using public key cryptography in such networks.  }
\RRmotcle{Réseaux Ad Hoc, protocoles cryptographiques, Diffie-Hellmann protocol}
\RRkeyword{Ad Hoc Networks, cryptographic protocoles, Diffie-Hellmann protocol}
\RRprojets{Codes et HiperCom}
\RRtheme{\THCom  \THSym } 
\URRocq  
\begin{document}
\makeRR   

\section{Introduction}

\par A Mobile Ad hoc NETwork (MANET) is a collection of mobile nodes connected
via a wireless medium forming an arbitrary topology. Implicit
herein is the ability for the network topology to change over time
as links in the network appear and disappear. To maintain the network
connectivity, a routing protocol must be used.  An important security
issue is that of the integrity of the network itself. Quite a lot of
studies have been already done to resolve security issues in existing routing
protocols (see \cite{ariadne},\cite{mae},\cite{securingolsr},\cite{rr5494}).

\par An orthogonal security issue is that of maintaining confidentiality
and integrity of data exchanged between nodes in the network. The task
of ensuring end-to-end security of data communications in MANETs is
equivalent to that of securing end-to-end security in traditional
wired networks. Many studies have been carried out to solve this
problem. One widespread solution is to create a virtual private
network (VPN) in a tunnel between the two communicating nodes. IPSec
is a well known security architecture which allows such VPNs to be
built between two communicating nodes.  However this solution requires a different secret key for each end-to-end connection.  Moreover the VPN solution can simply handle
unicast traffic.  An alternative solution is the use of a
shared secret key. There are many issues with such an approach. First
this key must be distributed among the network nodes. Second, to avoid
the compromising of this key it is required to renew the key often. A solution
to these two issues is the use a Group Key Agreement protocol, which
relies on the principles of the public key cryptography.

A Group Key Agreement protocol (GKA) is a key establishment technique in
which a shared secret is derived by more than two participants as a
function of information publicly contributed by each of them. They
are especially well suited to moderate sized groups with no central
authority to distribute keys. An authenticated group key agreement
protocol provides the property of key authentication (also called implicit
key authentication), whereby each participant is assured that no other
party besides the participants can gain access to the computed key. GKA
protocols are different from group key distribution (or key transport)
protocols wherein one participant chooses the group key and communicates
it to all others. GKA protocols help in deriving keys which are composed
of each one's contribution. This ensures that the resulting key is fresh
(for a given session) and is not favorable to one participant in any way.
The following security goals can be identified for any GKA protocol.
\begin{itemize} 
\item[1)]\textbf{Key Secrecy}: The key can be computed
only by the participants.  
\item[2)]\textbf{Key Independence}: Knowledge
of any set of group keys does not lead to the knowledge of any other
group key not in this set (see \cite{boyd-book}). 
\item[3)]\textbf{Forward Secrecy}:
Knowledge of some long term secret does not lead to the knowledge of
past group keys.  \end{itemize}

An important advantage of a group key agreement protocol over a simple
group key distribution scheme is the forward secrecy. This property can be 
particularly interesting in situations where some nodes are likely to be 
compromised (e.g. in military scenarios). In such scenarios, using a GKA, the knowledge of 
the long term secret of this node does not compromise all past session keys. 
From a functional point of view, it is desirable to have procedures to handle 
the dynamism in the network. These procedures enable efficient merging or 
partitioning of two groups in the network.

\section{Related Work} 
Key establishment protocols for networks can be broadly classified into 
three classes: \emph{Key transport using symmetric cryptography}, 
\emph{Key transport using asymmetric cryptography} and \emph{Key 
agreement using asymmetric cryptography}. In key transport protocols, 
one participant chooses the group key and securely transfers it to other 
participants using a priori shared secrets (symmetric or asymmetric). 
These protocols are not suitable for ad hoc networks for two reasons; 
firstly, they require a single trusted authority to distribute keys and 
secondly, compromise of the a priori secret of any participant breaches 
the security of all past group keys, thus failing to provide forward secrecy. 
Thus GKA protocols are more relevant since they provide this forward secrecy 
property. 

Most group key agreement protocols are derived from the two-party
Diffie-Hellman key exchange protocol. GKA protocols, not based on
Diffie-Hellman, are few and include the protocols of 
Pieprzyk and Li~\cite{li}, Tzeng and Tzeng~\cite{tzeng} and Boyd and Nieto ~\cite{boyd}. Both
protocols of Pieprzyk and Li~\cite{li} and Boyd and Nieto~\cite{boyd} fail to provide
\emph{forward secrecy} while the protocol of  Tzeng and Tzeng~\cite{tzeng} is quite
resource-intensive and prone to certain attacks \cite{boyd}. Forward
Secrecy is a very desirable property for key establishment protocols
in ad hoc networks, as some nodes can be easily compromised due to low
physical security of nodes. Thus it is essential that compromise of
one single node does not compromise all past session keys. We
summarize and compare in Table \ref{tb:gka-comp} existing GKA
protocols based on Diffie-Hellman protocols. We compare essentially
the unauthenticated versions of the protocols, as most achieve
authentication by using digital signatures in a very similar manner 
and thus have similar added costs for achieving authentication. 
We compare the efficiency of these protocols based on the
following parameters:
\begin{itemize}
\item \textbf{Number of synchronous rounds}: In a single synchronous
  round, multiple independent messages can be sent in the network. The
  total time required to run a round-efficient GKA protocol can be
  much less than other GKA protocols that have the same number of
  total messages but more rounds. This is because the nodes spend less
  time waiting for other messages before sending their own.
\item \textbf{Number of messages:} This is the total number of
  messages (unicast or broadcast) exchanged in the network to derive
  the group key. For multiple hop ad hoc networks, the distinction
  between unicast and broadcast messages is important as the latter
  can be much more energy consuming (for the whole network) than the
  former.
\item \textbf{Number of exponentiations}: All Diffie-Hellman based GKA
  protocols require a number of modular exponentiations to be
  performed by each participant. Relative to all cryptographic
  operations, a modular operation is the most computationally
  intensive operation and thus gives a good indication of the
  computational cost for each node.
\end{itemize}

Communication costs still remain the critical factor for choosing
energy-efficient protocols for most ad hoc networks. A modular
exponentiation (which is most efficiently done using elliptic curve cryptography) 
can be performed in a few
tens of milliseconds on most palmtops, whereas message propagation in
multi-hop ad hoc networks can be easily of the order of few seconds
and has energy implications for multiple nodes in the network. As can
be seen in Table \ref{tb:gka-comp}, most existing GKA protocols
require $O(m)$ rounds of communication for $m$ participants in the
protocol. Such protocols do not scale well in ad hoc networks. Even
tree-based GKA protocols with $O(\log m)$ rounds can be quite
demanding for medium to large sized ad hoc networks. Therefore
constant-round protocols are better suited for ad hoc networks.
\begin{table}
\begin{center}
\begin{tabular}{|c|c|c|c|c|}
\hline & Expo per $U_i$ & Messages & Broadcasts & Rounds\\ \hline
ITW \cite{itw} & $m$ & $m(m-1)$ & 0 & $m-1$ \\ 
GDH.1 \cite{gdh} & $i+1$ & $2(m-1)$ & 0 & $2(m-1)$\\ 
GDH.2 \cite{gdh, point3} & $i+1$ & $m-1$ & 1 & $m$\\ 
GDH.3 \cite{gdh} & 3 & $2m-3$ & 2 & $m+1$\\
Perrig \cite{perrig} & $\log_2 m +1$ & $m$ & $m-2$ & $\log_2 m$\\ 
Dutta  \cite{dutta} & $\log_3 m$ & $m$ & $m$ & $\log_3 m$\\ \hline
\end{tabular}\\
\end{center}
\caption{Comparison of non constant rounds GKA protocols} \label{tb:gka-comp}
\end{table}

\begin{table}
\begin{center}
\begin{tabular}{|c|c|c|c|c|c|c|}
\hline & Expo per $U_i$ & Messages & Broadcasts & Rounds & Structure & FS\\ \hline
Octopus \cite{octopus} & 4 & $3m-4$ & 0 & 4 & Hypercube & Yes\\ 
BDB \cite{bd, yung} & 3 & $2m$  & $m$ & 2 & Ring & Yes\\ 
BCEP \cite{bcep} & $2^{\dagger}$ & $2m$ & 0 & 2 & None & No\\ 
Catalano \cite{catalano} & $m+1$ & $2m$ & 0 & 2 & None & Yes\\ 
KLL \cite{kim} & $3$ & $2m$ &  $2m$ & 2 & Ring & Yes\\ 
NKYW \cite{won} &  $2^{\ddagger}$ & $m$ & $1$ & $2$ & None & Yes\\ 
STR \cite{ssdw,str} & $(m-i)^{\ast}$ & $m$ & $1$ & 2 & Skewed tree& Yes \\ 
Ours (AGDH) & $2^{\ast\ast}$ & $m$ & $1$ & 2 & None & Yes \\ \hline
\end{tabular}\\
\end{center}
$\dagger$: $m$ exponentiations for the base station.\\
$\ddagger$: $m+1$ exponentiations and m-1 inverse calculations for the parent node.\\
$\ast$: Up to $2m$ exponentiations for the sponsor node.\\
$\ast\ast$: $m$ exponentiations for the leader.
\caption{Comparison of constant round GKA protocols} \label{tb:gka-comp-cr}
\end{table}

Among the constant round protocols (see Table \ref{tb:gka-comp-cr}),
Octopus \cite{octopus}, BDB \cite{yung} and KLL \cite{kim} require
special ordering of the participants. This results in messages sent by
some participant being dependent on that of others. In such a case,
failure of a single node can often halt the protocol. Thus such
protocols are not robust enough to adapt well to the dynamism of ad
hoc networks. The BCEP protocol \cite{bcep} involves a base station,
and fails to provide forward secrecy if the long-term secret of the
base station is revealed. The Bresson and Catalano protocol
\cite{catalano} is computationally demanding with $O(m)$
exponentiations for each participant. Another drawback is that if any
participant's message is lost in first round, the whole protocol is
brought to a halt, as the secret sharing schemes implies all $m$
contributions are required to compute the key. Thus only the protocols
NKYW and STR (described below in details) seem to be usable in MANETs.

\textbf{NKYW}\cite{won}: The original paper proposes this protocol for ad hoc networks composed of devices 
with unequal computational powers. In the first round, each participant $M_i$ unicasts its 
contribution $g^{r_i}, i \in [1, n-1]$ to a fixed node $M_n$, called the parent node. The parent node 
chooses random $r$ and $r_n$ and 
computes $w = g^r$, $x_n = g^{rr_n}$ and $x_i$ = $(g^{r_i})^r$ for each received $g^{r_i}$. It broadcasts $w$ 
and \{$x_n * \Pi_{j \neq i} x_j $\}$_i$. 
The key is derived from $\Pi_i x_i$. The protocol remains a bit expensive computationally compared to 
the protocol that will be described in this paper.

\begin{figure}
\center { \setlength{\unitlength}{3947sp}%
\begingroup\makeatletter\ifx\SetFigFont\undefined%
\gdef\SetFigFont#1#2#3#4#5{%
  \reset@font\fontsize{#1}{#2pt}%
  \fontfamily{#3}\fontseries{#4}\fontshape{#5}%
  \selectfont}%
\fi\endgroup%
\begin{picture}(4216,3751)(2693,-4868)
{\color[rgb]{0,0,0}\thinlines
\put(5701,-1861){\circle{600}}
}%
{\color[rgb]{0,0,0}\put(6601,-2761){\circle{600}}
}%
{\color[rgb]{0,0,0}\put(4801,-2761){\circle{600}}
}%
{\color[rgb]{0,0,0}\put(5701,-3661){\circle{600}}
}%
{\color[rgb]{0,0,0}\put(4801,-4561){\circle{600}}
}%
{\color[rgb]{0,0,0}\put(3001,-4561){\circle{600}}
}%
{\color[rgb]{0,0,0}\put(3901,-3661){\circle{600}}
}%
{\color[rgb]{0,0,0}\put(5701,-2161){\line( 1,-1){600}}
}%
{\color[rgb]{0,0,0}\put(4801,-3061){\line( 1,-1){600}}
}%
{\color[rgb]{0,0,0}\put(3901,-3961){\line( 1,-1){600}}
}%
{\color[rgb]{0,0,0}\put(5701,-2161){\line(-1,-1){600}}
}%
{\color[rgb]{0,0,0}\put(4801,-3061){\line(-1,-1){600}}
}%
{\color[rgb]{0,0,0}\put(3901,-3961){\line(-1,-1){600}}
}%
\put(3001,-4561){\makebox(0,0)[lb]{\smash{{\SetFigFont{12}{14.4}{\rmdefault}{\mddefault}{\updefault}{\color[rgb]{0,0,0}$M_1$}%
}}}}
\put(4801,-4561){\makebox(0,0)[lb]{\smash{{\SetFigFont{12}{14.4}{\rmdefault}{\mddefault}{\updefault}{\color[rgb]{0,0,0}$M_2$}%
}}}}
\put(5701,-3661){\makebox(0,0)[lb]{\smash{{\SetFigFont{12}{14.4}{\rmdefault}{\mddefault}{\updefault}{\color[rgb]{0,0,0}$M_3$}%
}}}}
\put(6601,-2761){\makebox(0,0)[lb]{\smash{{\SetFigFont{12}{14.4}{\rmdefault}{\mddefault}{\updefault}{\color[rgb]{0,0,0}$M_4$}%
}}}}
\put(3001,-3961){\makebox(0,0)[lb]{\smash{{\SetFigFont{12}{14.4}{\rmdefault}{\mddefault}{\updefault}{\color[rgb]{0,0,0}$g^{r_1}$}%
}}}}
\put(4801,-3961){\makebox(0,0)[lb]{\smash{{\SetFigFont{12}{14.4}{\rmdefault}{\mddefault}{\updefault}{\color[rgb]{0,0,0}$g^{r_2}$}%
}}}}
\put(5701,-3061){\makebox(0,0)[lb]{\smash{{\SetFigFont{12}{14.4}{\rmdefault}{\mddefault}{\updefault}{\color[rgb]{0,0,0}$g^{r_3}$}%
}}}}
\put(6601,-2161){\makebox(0,0)[lb]{\smash{{\SetFigFont{12}{14.4}{\rmdefault}{\mddefault}{\updefault}{\color[rgb]{0,0,0}gr4}%
}}}}
\put(3901,-3061){\makebox(0,0)[lb]{\smash{{\SetFigFont{12}{14.4}{\rmdefault}{\mddefault}{\updefault}{\color[rgb]{0,0,0}$g^{r_1r_2}$}%
}}}}
\put(4801,-2161){\makebox(0,0)[lb]{\smash{{\SetFigFont{12}{14.4}{\rmdefault}{\mddefault}{\updefault}{\color[rgb]{0,0,0}$g^{r_3g^{r_1r_2}}$}%
}}}}
\put(5701,-1261){\makebox(0,0)[lb]{\smash{{\SetFigFont{12}{14.4}{\rmdefault}{\mddefault}{\updefault}{\color[rgb]{0,0,0}$g^{r_4g^{r_3g^{r_1r_2}}}$}%
}}}}
\end{picture}%
 } 
\caption{The STR Protocol}\label{fig:str}
\end{figure}
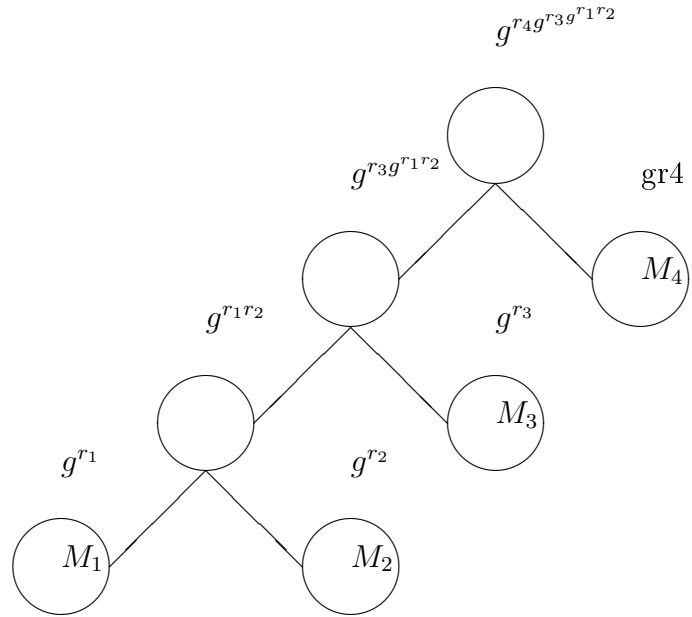

\textbf{STR}\cite{ssdw, str}: This protocol was proposed by Steer \textit{et al.} in \cite{ssdw} 
for static groups. Perrig \textit{et al.} proposed procedures to handle group changes in \cite{str}. 
Although this protocol has not been cited as a constant round protocol till now, we explain here in details 
why this protocol is indeed a constant round protocol. In the first round, each participant $M_i$ broadcasts 
its contribution $g^{r_i}$ (also known as its blinded key). In the second round, a key-tree as shown in 
Figure \ref{fig:str} where each leaf node represents a participant is constructed 
using participant IDs or the value of the contributions. The node 
in the bottom-most, left-most position in the tree is called the sponsor. The sponsor node broadcasts 
the set of blinded keys for all the intermediate nodes upto the root node. For the case shown in Figure \ref{fig:str}, the broadcast message is $\{ g^{r_1}, g^{r_2}, g^{r_3}, g^{r_4}$, $g^{g^{r_1r_2}}$, $g^{g^{r_3.g^{r_1r_2}}}\}$. The group key is $K$= $g^{r_4.g^{r_3.g^{r_1r_2}}}$. Participant $M_i$ has to perform $m-i$ exponentiations except the sponsor which has to compute $2m$ exponentiations. The protocol lacks a proof of security against active adversaries. 

Thus both these protocols are computationally more expensive compared to the protocol that will be 
described in this paper.

The contributions of this paper are the following:
\begin{itemize} 
\item an authenticated dynamic group key agreement protocol is
  recalled~\cite{tspuc},
\item the mechanisms that must be used in a MANET to implement
this group key agreement protocol are described, 
\item a precise study of the
cryptographic parameters that this group key agreement protocol must
use in the context of an ad hoc network is carried out.  
\end{itemize}

Finally the adapted version of the group key agreement protocol that we propose, we call 
this protocol AGDH for Asymetric Group Diffie Hellman, is among the very 
few group key agreement protocols suitable for ad hoc networks. 

The paper is organized as follows: 
\begin{itemize} 
\item Section ~\ref{sec-prespro} recalls the group key agreement  protocol. We describe the basic functioning of the protocol only, 
\item Section ~\ref{sec-usi}
explains how this group key agreement protocol can be implemented in an
ad hoc network. The main issues discussed in this section include the
election of a leader in the ad hoc network and the actions that must
be undertaken to handle splits and mergers in the ad hoc network,
\item Section ~\ref{sec-tun} discusses the overhead of cryptographic operations. 
\end{itemize}

\section{Presentation of AGDH} 
\label{sec-prespro}
We recall an existing group key agreement protocol in this section. We
first illustrate the basic principle of key exchange, followed by a
detailed explanation of how it is employed to derive Initial Key
Agreement, Join/Merge and Delete/Partition procedures to handle 
dynamism in ad hoc groups. 
\subsection{Notation} $G$: A subgroup (of prime order $q$ with generator
$g$) of some group.\\ 
$U_i$: $i^{th}$ participant amongst the $n$
participants in the current session.\\ 
$U_l$: The current group leader ($l
\in \{1,\dots,n\}$).\\ 
$r_i$: A random number (from $[1,q-1]$) generated by
participant $U_i$. Also called the \emph{secret} for $U_i$.\\ $g^{r_i}$:
The \emph{blinded secret} for $U_i$.\\ 
$g^{r_ir_l}$: The \emph{blinded
response} for $U_i$ from $U_l$.\\ 
$\mathcal M$: The set of indices of
participants (from ${\mathcal P}$) in the current session.\\ 
$\mathcal J$:
The set of indices of the joining participants.\\ 
$\mathcal D$: The set of
indices of the leaving participants.\\ 
$x \leftarrow y$: $x$ is assigned
$y$.\\ $x \stackrel{r} \leftarrow {\mathcal S}$: $x$ is randomly drawn
from the uniform distribution $S$.\\ 
$U_i \longrightarrow U_j: \{M\}$:
$U_i$ sends message $M$ to participant $U_j$.\\ 
$U_i \stackrel{B}
\longrightarrow {\mathcal M}: \{M\}$: $U_i$ broadcasts message $M$
to all participants indexed by $\mathcal M$.\\ 
$N_i$: Random nonce
generated by participant $U_i$.\\ 
${\mathcal V}_{PK_i}\{ msg_i, {\sigma}_i\}$: Signature verification algorithm which returns $1$ if $\sigma_i$ is a valid signature on message $msg_i$ else 0.
\subsection{A Three Round Protocol}

\subsubsection{The formal description}

Please note that in the following rounds each message is digitally
signed by the sender ($\sigma_i^j$ is signature on message $msg_i^j$
in Tables \ref{tb:1IKA}- \ref{tb:3DEL}) and is verified (along with
the nonces) by the receiver before following the protocol. Thus we omit to describe these 
steps which are formally shown in Tables \ref{tb:1IKA}- \ref{tb:3DEL}.

\textbf{Protocol Steps}:

\textbf{Round 1}: The chosen group
leader, $M_l$ makes a initial request ({\bf INIT}) with his identity,
$U_l$ and a random nonce $N_l$ to the group $\mathcal M$.

\textbf{Round 2}: Each interested $M_i$ responds to the {\bf INIT}
request, with a {\bf IREPLY} message which contains his identity $U_i$, a 
nonce $N_i$ and a blinded secret
$g^{r_i}$ to $M_l$ (see Table \ref{tb:1IKA} for exact message
contents).

\textbf{Round 3}: $M_l$ collects all the received blinded
secrets, raises each of them to its secret ($r_l$) and broadcasts them
along with the original contributions to the group, i.e. it sends
an  {\bf IGROUP} message that contains $\{U_i, N_i, g^{r_i}, 
g^{r_ir_l}\}$ for all $i \in {\mathcal M}
\setminus \{l\}$.

\textbf{Key Calculation}: Each $M_i$ checks if its contribution
is included correctly and obtains $g^{r_l}$ by computing
$(g^{r_ir_l})^{r_i^{-1}}$. The group key is \begin{displaymath}Key=
g^{r_l}* \Pi_{i \in {\mathcal M} \setminus \{l\}} g^{r_ir_l}\\ =
g^{r_l(1+\sum_{i \in {\mathcal M} \setminus \{l\}} r_i)}.\end{displaymath}

 \textbf{Note}:

 1) The original contributions $g^{r_i}$ are included in the last
 message as they are required for key calculation in case of group
 modifications (see below), and also, because it may be possible that
 a particular contribution has not been received by some member.

2) Even though $\Pi_{i \in {\mathcal M} \setminus \{l\}} g^{r_ir_l}$
is publicly known, it is included in key computation, to derive a key
composed of everyone's contribution. This ensures that the key can not be 
pre-determined and is unique to this session.

3) Even though the current group leader chooses his contribution after
others, he cannot pre-determine the group key.

The protocol is formally defined in Table~\ref{tb:1IKA}.
Table~\ref{tb:2JOIN} (respectively Table~\ref{tb:3DEL}) show how the
protocol is run when a group wants to join (respectively leave) an
existing group

\subsubsection{Example runs of the protocol}
We now see how this protocol can be used to derive Initial Key Agreement (IKA), Join/Merge and Delete/Partition procedures for ad hoc networks.

\begin{table}
\begin{center}
\begin{tabular}{|l|}\hline
\fbox{\textbf{Round 1}}\\
$l \stackrel{r} \leftarrow {\mathcal M}, N_l \stackrel{r} \leftarrow \{0, 1\}^k$\\
$U_{l} \stackrel{B} \longrightarrow {\mathcal M}:\{ msg_l^1=\{\mbox{ {\bf INIT} }, U_l, N_l\}, 
{\sigma}_l^1\}$\\
\fbox{\textbf{Round $2$}}\\
$\forall i \in {\mathcal M} \setminus \{l\}, if ({\mathcal V}_{PK_l}\{ msg_l^1, {\sigma}_l\}
 == 1)$, $r_i \stackrel{r} \leftarrow [1, q-1],  N_i \stackrel{r} \leftarrow \{0, 1\}^k$,\\ 
$U_i \longrightarrow U_{l}: \{ msg_i= \{ \mbox{ {\bf IREPLY}}  , U_l, N_l, U_i, N_i, g^{r_i}\}, 
{\sigma}_i\} $\\
\fbox{\textbf{Round $3$}}\\
$r_l \stackrel{r} \leftarrow [1, q-1]$,\\ 
$\forall i \in {\mathcal M} \setminus \{l\}, if ({\mathcal V}_{PK_i}\{ msg_i, {\sigma}_i\} == 1
)$ and $N_l$ is as contributed\\
$ U_{l} \stackrel{B} \longrightarrow {\mathcal M}: \{ msg_{l}^2 = \{\mbox {{\bf IGROUP}}, 
U_l, N_l, \{U_i, N_
i, g^{r_i}, g^{r_ir_l}\}_{i \in {\mathcal M} \setminus \{l\}}\}, {\sigma}_l^2\}$\\
\fbox{\textbf{Key Computation}}\\
$if ({\mathcal V}_{PK_l}\{ msg_l^2, {\sigma}_l^2\} == 1)$ and $g^{r_i}$ and $N_i$ are as contributed\\
$Key = g^{r_l(1+ \sum_{i \in {\mathcal M} \setminus \{l\}} r_i)}$\\
\hline
\end{tabular}
\caption{IKA}\label{tb:1IKA}
\end{center}
\end{table}

\paragraph{Initial Key Agreement} 

Secure ad hoc group formation procedures typically involve peer
discovery and connectivity checks before a group key is derived. Thus,
an $INIT$ request is issued by a participant and all interested peers
respond. The responses are collected and connectivity checks are
carried out to ensure that all participants can listen/broadcast to
the group (see for instance \cite{roman}). After the group membership
is defined, GKA procedures are implemented to derive a group key. Such
an approach is quite a drain on the limited resources of ad hoc
network devices. Thus an approach which integrates the two separate
procedures of group formation and group key agreement is required. The
above protocol fits well with this approach. Round 1 and Round 2 of
the above protocol can be incorporated into the group formation
procedures. In this way, blinded secrets, $g^{r_i}$'s, of all
potential members, $U_i$'s, are collected before the group composition
is defined. When the fully connected ad hoc group is defined, a single
broadcast message (Round 3 in Table \ref{tb:1IKA}) from the group leader, $U_l$,
(using contributions of only the joining participants) helps every
participant to compute the group key. An example is provided below.

Suppose $U_1$ initiates the group discovery and initially $5$ participants
express interest and send $g^{r_2}$, $g^{r_3}$, $g^{r_4}$, $g^{r_5}$ and
$g^{r_6}$ respectively along with their identities and nonces. Finally
only $3$ join because of the full-connectivity constraint. Suppose the
participants who finally join are $U_2$, $U_4$ and $U_5$. Then the group
leader, $U_1$, broadcasts the following message: \{$g^{r_2}$, $g^{r_4}$,
$g^{r_5}$, $(g^{r_2})^{r_1}$, $(g^{r_4})^{r_1}$, $(g^{r_5})^{r_1}$\}. On
receiving this message, each participant can derive $g^{r_1}$ using his
respective secret. Thus the key $g^{r_1(1+r_2+r_4+r_5)}$ can be computed.

\paragraph{Join/Merge}
\label{subsubsubJoinMerge}

\begin{table}
\begin{center}
\begin{tabular}{|l|}\hline
\fbox{\textbf{Round $1$}} \\
$\forall i \in {\mathcal J},r_i \stackrel{r} \leftarrow [1, q-1],  N_i \stackrel{r} \leftarrow 
\{0, 1\}^k$,\\ 
$U_i \stackrel{B} \longrightarrow {\mathcal M}: \{ msg_i= \{\mbox{ {\bf JOIN}}, U_i, N_i, g^{r_i}\}, {\sigma}
_i\}$\\  
\fbox{\textbf{Round $2$}}\\
$\forall i \in {\mathcal J}, if ({\mathcal V}_{PK_i}\{ msg_i, {\sigma}_i\} == 1) ~r_l \stackrel
{r} \leftarrow [1, q-1], l' \stackrel{r} \leftarrow {\mathcal M \cup \mathcal J}$\\
$U_l \longrightarrow U_{l'}: \{ msg_l= \{\mbox{ {\bf JREPLY}}, \{U_i, N_i, g^{r_i}\}_{\forall i \in {\mathcal
 M \cup \mathcal J}}\}, {\sigma}_l\}$\\
\fbox{\textbf{Round $3$}}\\
$if ({\mathcal V}_{PK_i}\{ msg_l, {\sigma}_l\} == 1), l \leftarrow l', r_l \stackrel{r} \leftarrow [1, q-1], {\mathcal M} \leftarrow {\mathcal M} \cup {\mathcal J}$\\
$ U_{l} \stackrel{B} \longrightarrow {\mathcal M}: \{ msg_{l}^2 = \{\mbox{ {\bf JGROUP}}, U_l, N_l, \{U_i, N_
i, g^{r_i}, g^{r_ir_l}\}_{i \in {\mathcal M} \setminus \{l\}}\}, {\sigma}_l^2\}$\\
\fbox{\textbf{Key Computation}}\\
$if ({\mathcal V}_{PK_l}\{ msg_l^2, {\sigma}_l^2\} == 1)$ and $g^{r_i}$ and $N_i$ are as contributed\\
$Key = g^{r_l(1+ \sum_{i \in {\mathcal M} \setminus \{l\}} r_i)}$\\
\hline
\end{tabular}
\caption{Join/Merge}\label{tb:2JOIN}
\end{center}
\end{table}

Suppose new participants, $U_9$ and $U_{10}$ join the group of
$U_1$, $U_2$, $U_4$ and $U_5$ with their contributions $g^{r_9}$ and
$g^{r_{10}}$ respectively. Then the previous group leader ($U_1$)
changes its secret to $r_1^{'}$ and sends $g^{r_1^{'}}$, $g^{r_2}$,
$g^{r_4}$, $g^{r_5}$, $g^{r_9}$, $g^{r_{10}}$ to $U_{10}$ (say the
new group leader). $U_{10}$ generates a new secret $r_{10}^{'}$ and
broadcasts the following message to the group: \{$g^{r_1^{'}}$, $g^{r_2}$,
$g^{r_4}$, $g^{r_5}$, $g^{r_9}$, $g^{r_{10}^{'}r_1^{'}}$, $g^{r_{10}^{'}r_2}$,
$g^{r_{10}^{'}r_4}$, $g^{r_{10}^{'}r_5}$, $g^{r_{10}^{'}r_9}$\}. And the new
key is $g^{r_{10}^{'}(1+r_1^{'}+r_2+r_4+r_5+r_9)}$.

\paragraph{Delete/Partition} 
\label{subsubsub-del}

When participants leave the group, they send a {\bf DEL} message, 
the group leader changes his secret
contribution and sends an \textbf{IKA} Round 3 like message to the
group, omitting the leaving participants' contributions. Refer to Table
\ref{tb:3DEL} and below for an example.

Suppose a participant, $U_2$, leaves the group of $U_1$, $U_2$, $U_4$,
$U_5$, $U_9$ and $U_{10}$. Then the leader, $U_{10}$ changes its secret
to $r_{10}^{''}$ and broadcasts $\{g^{r_1^{'}}$, $g^{r_4}$, $g^{r_5}$,
$g^{r_9}$, $(g^{r_1^{'}})^{r_{10}^{''}}$, $(g^{r_4})^{r_{10}^{''}}$,
$(g^{r_5})^{r_{10}^{''}}$, $(g^{r_9})^{r_{10}^{''}}\}$ to the group. And
the new key is $g^{r_{10}^{''}(1+r_1^{'}+r_4+r_5+r_9)}$.

\begin{table}
\begin{center}
\begin{tabular}{|l|}\hline
  \fbox{\textbf{Round $1$}}\\ $\forall i \in {\mathcal D}, U_i
  \longrightarrow U_l: \{ msg_i= \{\mbox{ {\bf DEL}}, U_i, N_i\}, {\sigma}_i\} $\\ 
  \fbox{\textbf{Round $2$}}\\ $\forall i \in {\mathcal D}, if
  ({\mathcal V}_{PK_i}\{ msg_i, {\sigma}_i\} == 1), r_l \stackrel {r}
  \leftarrow [1, q-1], {\mathcal M} \leftarrow {\mathcal M} \setminus
  {\mathcal D}$\\ $U_l \stackrel{B} \longrightarrow {\mathcal M}: \{
  msg_l = \{ \mbox{ {\bf DGROUP}}, U_l, N_l, \{U_i, N_i, g^{r _i}, g^{r_ir_l}\}_{i
    \in {\mathcal M} \setminus \{l\}}\}, {\sigma}_l\}$\\ 
  \fbox{\textbf{Key Computation}}\\ $if ({\mathcal V}_{PK_l}\{ msg_l,
  {\sigma}_l\} == 1)$ and $g^{r_i}$ and $N_i$ are as contribute d\\ 
  $Key = g^{r_l(1+ \sum_{i \in {\mathcal M} \setminus \{l\}} r_i)}$\\ 
  \hline
\end{tabular}
\caption{Delete/Partition}\label{tb:3DEL}
\end{center}
\end{table}

\section{Using this GKA protocol within an ad hoc network}
\label{sec-usi}

In the following we are considering a multi-hop ad hoc network. We are not
assuming any particular property of the routing protocol which ensures
the connectivity of the network. We can use reactive protocols as AODV or DSR 
\cite{aodv,dsr} where 
the connectivity is created on demand when a route is needed. We can also use 
proactive protocols as OLSR or TBRPF \cite{olsr,tbrpf} 
where synchronous packets are used to maintain the knowledge of the 
topology. We will assume that we have
a broadcast mechanism to flood messages within the ad hoc network. We 
are not assuming that this flooding mechanism is reliable, but
we assume that the network is connected and that flooding 
messages finally reaches all the network nodes \footnote{We mean that synchronous 
flooded messages will finally reach all the network nodes even if there 
are messages loosses}.

A key point in the GKA protocol described above is the existence of
group leader. Thus it is necessary to have a robust mechanism to elect
such a leader in an ad hoc network.  That is the first issue that we study.

\subsection{Election of a group leader} 
\label{sub-ele}

A key requirement is that all members of a group agree on the same
group leader. A simple solution is that the group leader periodically
broadcasts messages. These messages then serve as a proof, for nodes
that are within reach of the group leader, that a group leader exists and
operates properly.  We can simply use the \textbf{INIT} message of GKA protocol to
demonstrate the existence and the correct functioning of the group
leader. When the other nodes in the network receive this \textbf{INIT} message
each replies with an \textbf{IREPLY} message including their contribution. 
Using these \textbf{IREPLY} messages, the group leader defines a group and sends to all members of the group an \textbf{IGROUP} message. The \textbf{INIT} message 
can be seen as an \textbf{IGROUP} message when the group is not yet defined. In the following we will only use 
the term \textbf{IGROUP} message.

These \textbf{IGROUP} messages are sent periodically; depending 
on the dynamics of the group, the group leader will send a new \textbf{IGROUP} message or exactly
the same message as before. If the network only comprises of the group leader, the
latter will send periodically empty \textbf{IGROUP} messages. It will stop sending this message when a node joins its network by replying to
its \textbf{IGROUP} message with an \textbf{IREPLY} message. The mechanism to elect
a group leader simply follows from the property that, in a network with
a group leader, periodic messages are broadcasted by the group leader and are,
in principle, received by the group members. If a node does not receive
a message for a fixed period T, known a priori by the network nodes,
this node sets a random timer. At the expiration of this timer and
if no \textbf{IGROUP} message
has been received meanwhile, the node becomes the group leader. It
then sends an empty \textbf{IGROUP} message. 

There may be a collision on
\textbf{IGROUP} messages if two nodes or more have selected the same
value for their random timer. In such a case, there may be \textbf{IGROUP} 
messages generated by two (or more) group leaders. To
select a group leader, we can use additional rules. The first rule is
that when a group leader A receives an \textbf{IGROUP}
message from a group leader B which has a smaller ID than its own ID,
the group leader A just stops to send its periodic messages. The group
members that will receive periodic messages from more than one group
leader will only consider the message issued by the group leader with
the smallest index. Thus if an \textbf{IGROUP} message showing a larger
ID than a previously received \textbf{IGROUP} message is received, then this message
is simply discarded and no \textbf{IREPLY} message is issued. On the
contrary if an \textbf{IGROUP} message showing a smaller ID is received
then the node issues a \textbf{IREPLY} message.

\par
Another issue is how the GKA protocol takes into account the dynamism
of an ad hoc network. For instance a node may leave the network without
being able to send the group leader a message pointing out its departure
from the network. This issue is handled in the next subsection

\subsection{Handling join and withdrawal of a node}
\label{sub-join}

A node which joins the network will receive the periodic
\textbf{IGROUP} message of the group leader. He will just have to send
a \textbf{JREPLY} message, with its contribution, to join the group.
The group leader will incorporate this new contribution in its next
\textbf{IGROUP} message. Actually there is no need in the protocol to
differentiate between \textbf{JREPLY} and \textbf{IREPLY}.  Thus, for
simplicity sake, we will only keep the \textbf{IREPLY} message.

In an ad hoc network, the only conceivable way for the group leader to
be sure that a node still belongs to a group is to receive a message from
it. Thus to handle the dynamism of a group, the group leader will use
the periodic reception of the \textbf{IREPLY} messages.  The period
with which an \textbf{IREPLY} message is sent by a member of the group
should be the same for all the nodes of the group. If the group leader
is not receiving a \textbf{IREPLY} message for a given number of
periods (greater than 1 to handle possible packet loss), the lack of
reception of these messages should be handled in the same way as the
reception of a \textbf{DEL} message.  In such a case the group leader
will change its own contribution in the \textbf{IGROUP} message and
will re-send the \textbf{IGROUP} message.

When a node deliberately wishes to withdraw from a group it can use the 
\textbf{DEL} message to announce this wish to the group leader. 
Upon the reception of such a message the group leader will change its 
own contribution in the \textbf{IGROUP} 
message and will re-send the \textbf{IGROUP} message. The use of the 
\textbf{DEL} message will speed up the taking into account of the node 
withdrawal. 

\subsection{Handling merge or split of groups}

The merger of groups (two or more) leads group leaders to receive
\textbf{IGROUP} messages from other group leaders.  The scheme used in
the group leader election can be used to resolve the conflict.  When
the conflict is resolved only one group leader is left in the group.
If a group splits, a part of the group will remain without group
leader. The technique used in the group leader election can be used in
the subgroups without leader to elect a new leader.

\subsection{Renewing its contribution}

The group leader and group members will have to renew their contribution 
periodically. For the group leader, the change of its contribution or of some 
member of the group will lead to a change in the content of the \textbf{IGROUP} 
message. To simplify we can assume that the group leader and the group members change 
their contribution at the same rate. 

We have given all the principles of the protocol. We precise the details of the 
whole protocol in the next section.

\subsection{Implementation issues}

We will consider a given period $T$. To simplify, this period will be used both 
by the group leader or by the member of the group as a period to send their 
GKA messages. 

A node can be in one of the following two states : \textbf{member
  state} or \textbf{group leader state}.  A node in a member state
will enter the process to become a group leader if it has not received
\textbf{IGROUP} message for a duration $kT$.  A node which has not
received any message from a group leader for a duration $kT$ with $k
\ge 2$ will suppose that there is no group leader and starts the
procedure to become a leader.  Since a node may not have received a
packet of the group leader because this packet has been lost, $k$ must
be selected so that the probability that $k-1$ successive
transmissions of a GKA message are lost is small. Then, to become a group
leader, the node selects a random integer $i_r$ between 1 and a given
number $l$ (backoff window size) and initializes a timer at $i_r t_{rtd}$, where $ t_{rtd}$ is a
predefined duration computed to be at least the round trip delay of a
message throughout the ad hoc network. With such a figure for
$t_{rtd}$ we can be sure that if two nodes draw different integers
$i_r$ and $i_{r'}$, the node having selected the larger integer will
receive the \textbf{IGROUP} message of the other node and then will
stop its election process. The backoff window size $l$ must be chosen with respect to the
total number of nodes in the network so that the probability that two
nodes choose the same integer is small.  This back-off procedure is
performed to avoid possibly multiple group leader candidates, for
instance, when a group is set up or split into two subgroups.

\dessin{master-leader}{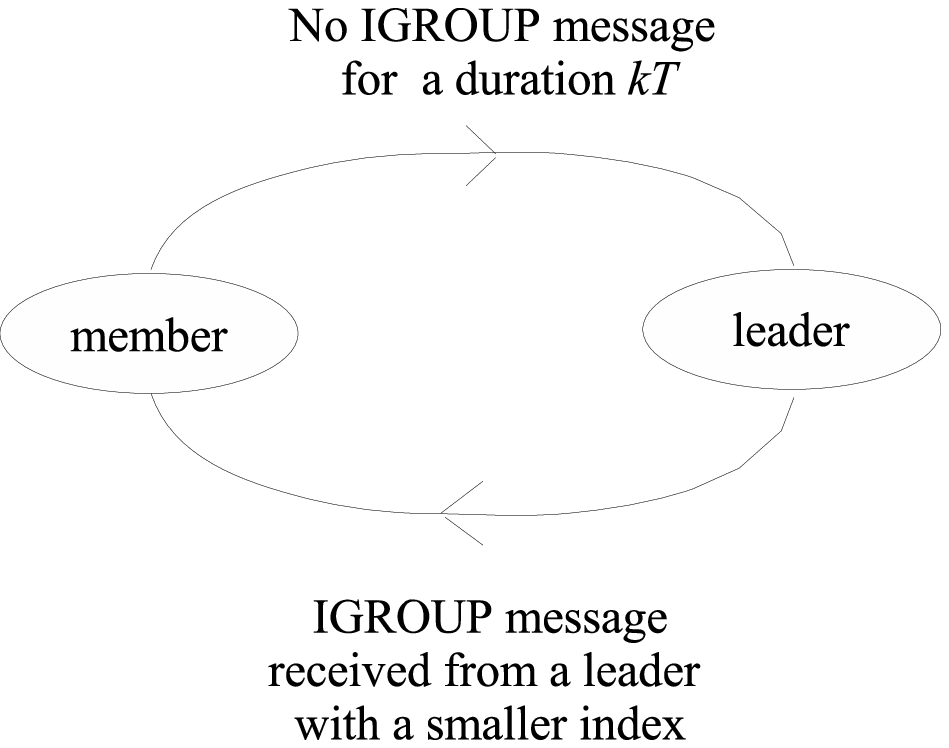}{7}{Transition between the
  member and the leader state}

When the node in the state member sends its first \textbf{IGROUP}
message, it is in the group leader state, see Figure
\ref{master-leader}.  In the group leader state, a node must collect
\textbf{IREPLY} messages and form the related \textbf{IGROUP} message.
When there is a change in the group (arrival or withdrawal) the group
leader must change its contribution. Additionally, irrespective of the
modification of the composition of the group, the group leader must
change its contribution periodically, to maintain the security of the session 
key. 

When a group leader is elected, the latter may choose to wait
additional periods before sending a \textbf{IGROUP} containing the
contributions of the group members. Doing so, the group leader may
avoid unnecessary changes to the session key due to the lack of
receipt of all contributions in time.

In the group leader state, a node will also look out for \textbf{IGROUP} 
messages from another group leader. If it receives such a message from another group 
leader holding a smaller node index, the node changes its state to the member state. 
In the member state, a node will have to send \textbf{IREPLY}
messages periodically. Like the group leader, a group member must
change its contribution periodically with a period $P$ see figure \ref{renew}. We will
assume that $P$ is a large multiple of $T$. To simplify the procedure
and to avoid unnecessary computations we can assume that the group
leader does not instantly include a new contribution of a group member
in the \textbf{IGROUP} message, instead it will wait for the change of
its own contribution to take into account all new contributions of
nodes. This is possible since the contribution of the node member is
included in the \textbf{IGROUP} message.

\dessin{renew}{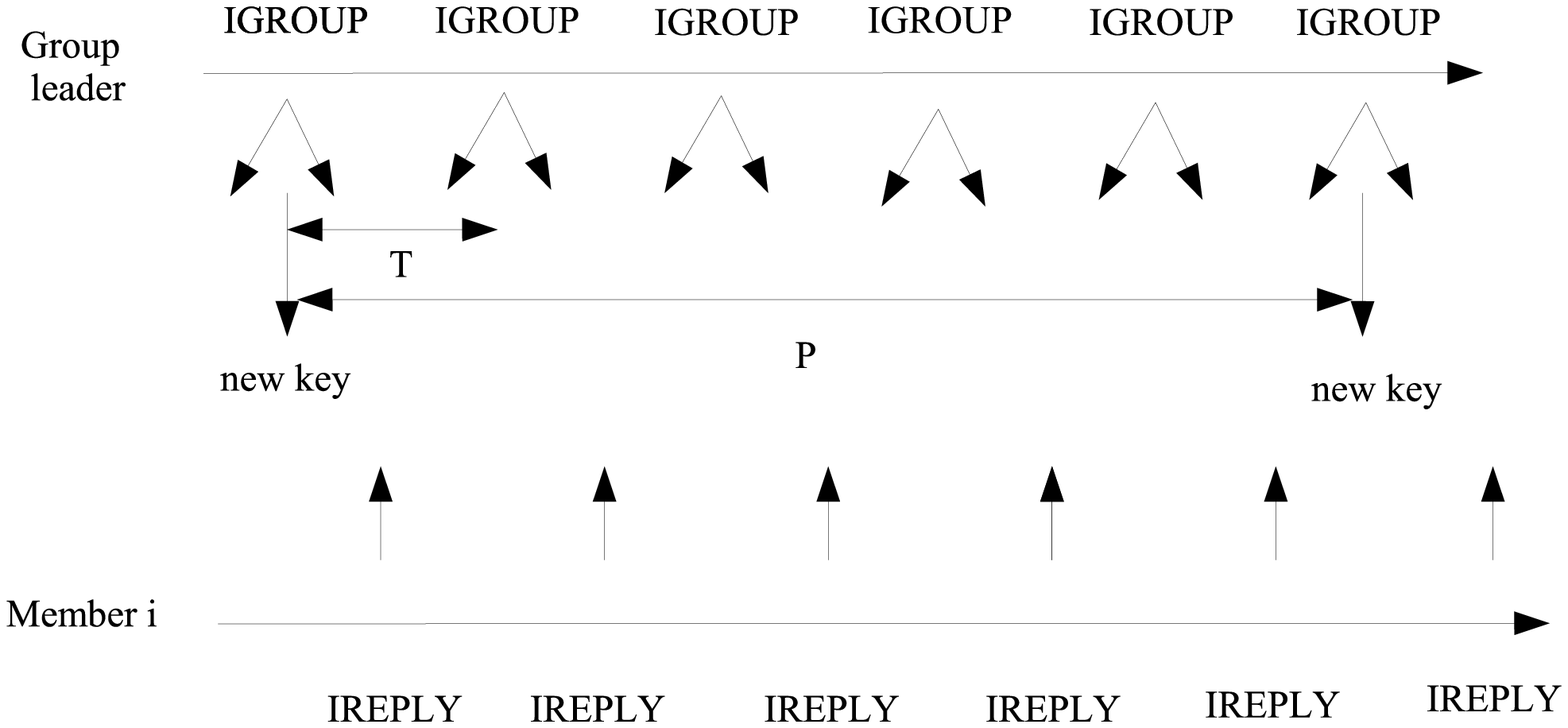}{10}{Sending \textbf{IGROUP} and  \textbf{IREPLY} messages}

Both \textbf{IGROUP} and \textbf{IREPLY} messages must be sent
periodically for each interval $T$. To reduce the probability of
collision of these messages, we add a jitter to times when the GKA
messages shall be sent by the group members and the group leader.

In the table \ref{parameters}, we have given examples of figures for our GKA protocol. 
We can notice 
that  $l$ and $t_{rtd}$ will heavily depend of the number of nodes in the network and 
of the topology of 
the network.  
\begin{center}
\begin{table} 
\begin{center}
\begin{tabular}{|c|c|p{1.7 in}|}\hline
Parameter  &    Value         &  Constraint     \\
\hline $P$: key renew period  &    20 min     &      \\ 
\hline $T$: period of &       & 
\\ IGROUP messages &   5s  &      \\ 
\hline $k$: number of messages  &    &  large enough to be sure that\\
losses before assuming    &    3          &   the message is not simply \\
a node leaves   &             &  lost \\
\hline $l$: backoff window   &   20          &   large enough to avoid collision during the group leader election\\ 
\hline $t_{rtd}$: backoff slot  for &         &  more than a round \\
leader election             &  100 ms &  trip delay  \\ 
\hline 
\end{tabular} 
\caption{Protocol parameters }
\label{parameters} 
\end{center}
\end{table}
\end{center}

\section{Computational overhead}
\label{sec-tun}

\begin{figure}
\begin{center}
\begin{tabular}{|l|c|c|c|}\hline
Group& Size of contributions &blindings/second=recoveries/second\\ \hline
Modular Field & 1024 bits &10\\\hline
Elliptic curve &160 bits&93\\\hline
\end{tabular}
\end{center}
\caption{Performance of elliptic curve cryptography, versus a classical group (modular integers)on a iPAQ, StrongARM-1110, using the {\tt openssl} implementation, for a security level of $2^{80}$. Blinding means computing
  $ g^{r_i}$, and recovering means computing $g^{r_0}$ from the
  blinded response $g^{r_ir_0}$ of the leader .}\label{fig:ecc_cost}
\end{figure}

Figure~\ref{fig:ecc_cost} describes the cost, on an average small device (COMPAQ iPAQ),
of elliptic curve
cryptography which is more efficient 
than classical cryptographic relying on biger groups.
Basically, for a security level of $2^{80}$, such a device can perform
almost 100 operations per second. Thus the latency of elliptic curve
exponentiation is 10 msec per device, except for the leader whose
computational cost grows linearly with the size of the group. Thus
there is concern for this particular node. Assuming that the leader
devotes half its times towards cryptogaphic operations, managing a
group of size 50 will impose a delay of 1 second before being able to
send the blinded response.

The above computational load on the group leader is in the case where
the group leader receives all the blinded secrets at once, and has to
give the blinded response also at once. In practice, the group leader
will receive the blinded secret at different time slots. It is then
possible to perform operations in batch: the group leader can
generates its own secret in advance, and compute on the fly the
blinded reponses $(g^{r_i})^{r_0}$ upon reception of each blinded
secret $g^{r_i}$. He can also stepwise compute the product
$(g^{r_1})^{r_0}\cdots (g^{r_m})^{r_0} $, where $m$ is the index of
the last received contribution. When he has to broadcat the {\tt
  IGROUP} message, all the computationaly intense cryptographical
operations, necessary to generate the blinded responses, have already
been performed.


\section{Conclusion}
\label{sec-conclusion}

We have discussed a group key agreement protocol for 
handling ad hoc group of small to moderate size. We have fully specified the implementation details 
needed for actual use of the protocol, relying on know network techniques such as 
self election, periodic broadcast, back-off techniques. 
The protocol is robust in the sense 
that connectivity losses does not impair its functioning. We have experienced that the 
computational cost of public key cryptography is kept reasonably low. 
If we consider constraints in ad hoc networks: no network structure, high dynamism, 
restricted bandwidth the presented protocol is among the few GKA protocols 
which is suitable for ad hoc networks. 

\newcommand{\etalchar}[1]{$^{#1}$}

\end{document}